\documentclass[10pt,article]{IEEEtran}

\usepackage[dvips]{graphicx}
\usepackage[letterpaper,pdftex,margin=0.625in]{geometry}
\usepackage{graphicx}
\usepackage{subfigure}
\usepackage{epsf}
\usepackage{amsmath,amsthm,amssymb}
\usepackage{multirow}

\newtheorem{definition}{Definition}
\newtheorem{example}{Example}

\newtheorem{thm}{Theorem}

\newtheorem{cons}{Construction}

\begin{document}

\title{Algebraic symmetries of generic $(m+1)$ dimensional periodic Costas arrays}


\author{Jos\'{e} Ortiz-Ubarri, Oscar Moreno, Andrew Z. Tirkel, Rafael Arce-Nazario, Solomon W. Golomb\thanks{J. Ortiz-Ubarri and R. Arce-Nazario are with the Computer Sciences Department, University of Puerto Rico, San Juan Puerto Rico, E-mails: jose.ortiz@hpcf.upr.edu and rafael.arce@upr.edu.; O. Moreno is with the Gauss Research Laboratory Inc., San Juan, Puerto Rico. E-mail: moreno@nic.pr.; A. Tirkel is with the Scientific Technology Pty Ltd, Australia. E-mail: atirkel@bigpond.net.au.; S.W. Golomb is with the University of Southern California, Los Angeles, California, US. Email: sgolomb@usc.edu.}}

\maketitle


\begin{abstract}
In this work we present two generators for the group of symmetries of the generic $(m+1)$ dimensional periodic Costas arrays over elementary abelian $(\mathbb{Z}_p)^m$ groups: one that is defined by multiplication on $m$ dimensions and the other by shear (addition) on $m$ dimensions.  Through exhaustive search we observe that these two generators characterize the group of symmetries for the examples we were able to compute.  Following the results, we conjecture that these generators characterize the group of symmetries of the generic $(m+1)$ dimensional periodic Costas arrays over elementary abelian $(\mathbb{Z}_p)^m$ groups.
\end{abstract}

\section{Introduction}

The original two dimensional Costas arrays were introduced in~\cite{Costas, golomb84algebraic} and their periodicity properties have been studied in~\cite{GolombTaylor, MorenoGolombCorrada}, and more recently in~\cite{MorenoOrtizITW2009}.  Good correlation, and full periodicity properties  such as those of some Costas arrays are very important for applications in digital watermarking~\cite{tirkelelectronic, TirkelHall-MorenoMaric}, and optical communications~\cite{KYP3DOOC,OrtizMorenoTirkel}.
 
Multidimensional periodic arrays are especially useful for applications of digital watermarking for video, and combined video and audio media. In recent work Moreno et al.~\cite{MTGK} presented a multi-periodical generalization of the Welch construction over the elementary abelian group $(\mathbb{Z}_p)^m$.  This Welch construction is too sparse for  the application of digital watermarking, but in their work they show how its multi-periodicity property allows the substitution of columns by periodic sequences with good correlation properties.  Three dimensional periodic arrays also have applications in optical communications.

In this work, we generate many multidimensional periodic Costas arrays (MCPA) over elementary abelian $(\mathbb{Z}_p)^m$ groups. These arrays are obtained through the application of algebraic symmetries to the generalization of the multidimensional periodic Welch construction.  We call these symmetries: \textbf{$G_1$} multiplication on the $m$ dimensions symmetries and \textbf{$G_2$} shear (addition) on the $m$ dimensions symmetries.  We independently found \textbf{$G_1$} and \textbf{$G_2$} analysing the results obtained through exhaustive exploration of some examples of these Welch arrays; later we found that these kinds of algebraic permutations were used on Quadratic residue arrays by B\"{o}mer et al~\cite{bomerQRA}, and on two dimensional Costas and sonar arrays~\cite{MorenoGamesTaylor}.  

We compared the number of results obtained from the exhaustive exploration of the examples $(\mathbb{Z}_3)^2$,  $(\mathbb{Z}_5)^2$, and  $(\mathbb{Z}_2)^3$ with the number of results obtained using these generators; and observed that the group of symmetries for the examples is characterized with the use of the algebraic generators.  In other words, all the MPCA found through exhaustive exploration are generated using a combination of the symmetries that will be presented in this work.

The rest of the paper is organized as follows:  Section \ref{sec:def} contains the formal definitions of a generic $(m+1)$-dimensional periodic Costas array over the elementary abelian group $(\mathbb{Z}_p)^m$, the definition of the Costas property, and reviews two algebraic symmetries of multidimensional periodic Welch Costas arrays that are used in combination with the symmetries presented in this work to characterize the group of symmetries of the generic MPCA. Sections \ref{sec:g1} and \ref{sec:g2} present the generators used to complete the characterization of the group of symmetries of the generic 3D MPCA. Sections \ref{sec:g1g} and \ref{sec:g2g} present the generalization of the generators for multiple dimensions.  Section \ref{sec:comp} describes the computational work performed; and section \ref{sec:con}  concludes this work, with our conjectures on the existence of the generic multidimensional periodic Costas arrays.

\subsection{Definitions}\label{sec:def}

We begin by providing the definitions of the generic and general $(m+1)$-dimensional periodic Costas array over the elementary abelian group and introduce the Generalized Welch construction.

\begin{definition}
 \emph{A generic $(m+1)$-dimensional periodic Costas array over the elementary abelian group $(\mathbb{Z}_p)^m$} is a permutation function $f: ((\mathbb{Z}_p)^m)^* \rightarrow (\mathbb{Z}_{p^m -1})$, where $A^*$ means A-\{0\}. This function has the \emph{distinct difference property}: for any $h \ne 0, a, b \in (\mathbb{Z}_p)^m, f(a + h) - f(a) = f(b+h) - f(b)$ implies $a = b$, where the addition and subtraction operations are performed in the corresponding abelian group. 
\end{definition}

\begin{example}\label{ex:3DW}
The following is a grid defined over $\mathbb{Z}_5 \times \mathbb{Z}_5$:

\[ \textbf{W} =\left( \begin{array}{ c c c c c }
(4,0) & (4,1) & (4,2) & (4,3) & (4,4) \\
(3,0) & (3,1) & (3,2) & (3,3)& (3,4) \\
(2,0) & (2,1) & (2,2) & (2,3) & (2,4) \\
(1,0) & (1,1) & (1,2) & (1,3) & (1,4) \\
(0,0) & (0,1) & (0,2) & (0,3) & (0,4) \end{array} \right)\]

The following is a three dimensional periodic Welch Costas array over the elementary abelian group $\mathbb{Z}_5 \times \mathbb{Z}_5$:

\[ \textbf{W} =\left( \begin{array}{ c c c c c }
1 & 17 & 14 & 15 & 10 \\
7 & 21 & 23 & 16 & 20 \\
19 & 8 & 4 & 11 & 9 \\
13 & 22 & 3 & 2 & 5 \\
$*$ & 0 & 6 & 18 & 12 \end{array} \right)\]
\end{example}

\textbf{Remark:} Through the examples of this work the rows are numbered upwards, starting from the bottom, whilst the columns are numbered from left to right.

\begin{definition}
\emph{A general $(m+1)$ periodic Costas array} is defined as before, but $*$ is now over any grid location.
\end{definition}

The well known Welch construction is generalized to construct MPCA.

\begin{cons}
 \textbf{(Generalized Welch:  a multi-periodic Costas array in $m+1$ dimensions with correlation 1).}\\
 
 Consider the sequence $s$ such that $s_{\alpha^i} = log_\alpha(X)$ where $X \in GF(p^m)$, and $\alpha$ is a primitive element of $GF(p^m)$. Specifically for $X = \alpha^i$, $s_{\alpha^i} = log_\alpha \alpha^i = i$.  The sequence scheme is to use the m-tuple representation of $\alpha^i$ to determine the coordinates (location) on the m dimensional integer grid defined above.  $s$ is a periodic sequence with period $p^m - 1$.  For our Costas type construction we take the grid point location belonging to $s_{\alpha^i}$, and place a 1 in a column of length $p^m-1$ located above the grid point, and zeros in all other entries in that column.
\end{cons}

\begin{thm}
The permutation $s$ posseses the Costas property:
$s_{(a+h)} - s_{a} = s_{(b+h)} - s_{b} \rightarrow a = b$, $\forall h \ne 0$  and is therefore a Costas array.
\end{thm}

\begin{IEEEproof}

Let $a, b, h \in GF(p^m)$ and $h \ne 0$

\begin{eqnarray}
log(a+h) - log(a) & = & log(b+h) - log(b)\\
log\left(\frac{a+h}{a}\right) & = & log\left(\frac{b+h}{b}\right)\\
1 + \frac{h}{a} & = & 1 + \frac{h}{b}  
\end{eqnarray}

Therefore, for $h \ne 0 \rightarrow a = b$.

\end{IEEEproof}

\subsection{Addition and Multiplication (modulo  $p^m-1$) symmetries}

The following two algebraic symmetries were introduced by Moreno et al. in a previous work~\cite{MTK} and are used to characterize the group of symmetries of our examples. The multiplication (modulo $p^m-1$) is the first symmetry applied to the sequences obtained from the Welch construction, then we apply the \textbf {G1} and \textbf{G2} symmetries introduced in this work, and finally we use the addition symmetry to finish the characterization of our examples.

\begin{thm}\label{mthm2}
Multiplication (modulo $p^m-1$) of a periodic Costas array by an integer less than and relatively prime to $p^m-1$ generates a new periodic Costas array.
\end{thm}

\begin{example}\label{mex2} Multiplication of \textbf{W} by $23\equiv -1\bmod 24$.

\[\left( \begin{array}{ c c c c c }
23 & 7 & 10 & 9 & 14\\
17 & 3 & 1 & 8 & 4\\
5 & 16 & 20 & 13 & 15\\
11 & 2 & 21 & 22 & 19\\
$*$ & 0 & 18 & 6 & 12\\
\end{array}\right)\]
\end{example}

\begin{thm}\label{mthm3} Addition (modulo $p^m-1$) of any integer less than $p^m-1$ to a periodic Costas array generates a new periodic Costas array. \end{thm}

\begin{example}\label{mex3} Addition of 4 to \textbf{W}.

\[\left(\begin{array}{ c c c c c }
5 & 21 & 18 & 19 & 14\\
11 & 1 & 3 & 20 & 0\\
23 & 12 & 8 & 15 & 13\\
17 & 2 & 7 & 6 & 9\\
$*$ & 4 & 10 & 22 & 16\\
\end{array}\right)\]
\end{example}

The addition and multiplication (modulo $p^m-1$) symmetries are used in combination with the symmetries presented in the next sections to generate many MPCA and to characterize the group of symmetries of the examples that we also characterized computationally.

\section {Row and Column Multiplication}\label{sec:g1}

For the 3D Welch MPCA, the rows and columns multiplication symmetries consist in shifting the rows and columns of the two dimensional representation.  In general, these symmetries consists of shifting m-dimensions of the $(m+1)$ dimension Welch MPCA $x$ times their original position.  In the following sections we define the row multiplication (RG1), column multiplication (CG1) and the combination of the row and column multiplication symmetries (G1), and apply them to the 3D Welch MPCA.  Finally we define a generalization of the \textbf{G1} symmetry for $(m+1)$ dimensions.

\subsection{Row Multiplication}\label{subsec:rm}

Let $(i, j) \in (\mathbb{Z}_p)^2$, and $x \in \{1 \dots p-1\}$, and let $(\textbf{w}_{i, j})$ be the element of matrix \textbf{W} in position $(i,j)$.

\begin{definition}(Row Multiplication)\\
$\textbf{RG1}_x: (\mathbb{Z}_p)^2 \rightarrow (\mathbb{Z}_p)^2$ is a permutation function such that $\textbf{RG1}_x:(\textbf{w}_{i, j}) \rightarrow (\textbf{w}_{xi , j})$ permutes the rows from position $i$ to position $xi$ of a three dimensional periodic Costas \textbf{W} over the elementary abelian group $\mathbb{Z}_p \times \mathbb{Z}_p$. 
\end{definition}

\begin{example}

The following three dimensional periodic Welch Costas array over the elementary abelian group $\mathbb{Z}_5  \times  \mathbb{Z}_5$ is one symmetry, out of $4$ distinct row multiplications symmetries obtained after applying $\textbf{RG1}_2(\textbf{W})$. 

\[\textbf{RG1}_2(\textbf{W}) = \left( \begin{array}{ c c c c c }
19 & 8 & 4 & 11 & 9 \\
1 & 17 & 14 & 15 & 10 \\
13 & 22 & 3 & 2 & 5 \\
7 & 21 & 23 & 16 & 20 \\
$*$ & 0 & 6 & 18 & 12 \end{array} \right)\]
\end{example}

\subsection{Column Multiplication}\label{subsec:cm}

Let $(i, j) \in (\mathbb{Z}_p)^2$, and $x \in \{1 \dots p-1\}$, and let $(\textbf{w}_{i, j})$ be the element of matrix \textbf{W} in position $(i,j)$.

\begin{definition}(Column Multiplication) \\
$\textbf{CG1}_x: (\mathbb{Z}_p)^2 \rightarrow  (\mathbb{Z}_p)^2$ is a permutation function such that $\textbf{CG1}_x:(\textbf{w}_{i, j}) \rightarrow (\textbf{w}_{i , xj})$ permutes the columns from position $j$ to position $xj$ of a three dimensional periodic Costas \textbf{W} over the elementary abelian group $\mathbb{Z}_p \times \mathbb{Z}_p$.
\end{definition} 

\begin{example}

The following three dimensional periodic Welch Costas array over the elementary abelian group $\mathbb{Z}_5  \times  \mathbb{Z}_5$ is one symmetry, out of $4$ distinct column multiplications symmetries obtained after applying the $\textbf{CG1}_3(\textbf{W})$. 

\[\textbf{RG1}_3(\textbf{W}) = \left( \begin{array}{ c c c c c }
1 & 14 & 10 & 17 & 15 \\
7 & 23 & 20 & 21 & 16 \\
19 & 4 & 9 & 8 & 11 \\
13 & 3 & 5 & 22 & 2 \\
$*$ & 6 & 12 & 0 & 18 \end{array} \right)
\]
\end{example}

We can combine both transformations to generate more Welch Costas arrays.

Note that $\textbf{RG1}_{x_1}(\textbf{CG1}_{x_2}(\textbf{W})) = \textbf{CG1}_{x_2}(\textbf{RG1}_{x_1}(\textbf{W}))$.

\begin{example}

The following is the three dimensional periodic Welch Costas array over the elementary abelian group $\mathbb{Z}_5  \times  \mathbb{Z}_5$ after applying the $\textbf{RG1}_{2}(\textbf{CG1}_{3}(\textbf{W}))$. 

\[ \textbf{RG1}_{2}(\textbf{CG1}_{3}(\textbf{W})) =\left( \begin{array}{ c c c c c }
19 & 4 & 9 & 8 & 11 \\
1 & 14 & 10 & 17 & 15 \\
13 & 3 & 5 & 22 & 2 \\
7 & 23 & 20 & 21 & 16 \\
$*$ & 6 & 12 & 0 & 18 \end{array} \right)\]

\end{example}

\subsection{Row and Column Multiplication}\label{subsec:rcm}
\begin{definition}(Row and Column Multiplication)
$\textbf{G1}_{x_1,x_2} :(\mathbb{Z}_p)^2 \rightarrow  (\mathbb{Z}_p)^2$ is a permutation function such that $\textbf{G1}_{x_1,x_2}:(\textbf{w}_{i, j}) \rightarrow (\textbf{w}_{x_1i , x_2j})$ permutes the rows from position $i$ to position $x_1i$ and the columns from position $j$ to position $x_2j$ of a three dimensional periodic Costas \textbf{W} over the elementary abelian group $\mathbb{Z}_p \times \mathbb{Z}_p$.
\end{definition} 

\begin{example}

The following is the three dimensional periodic Welch Costas array over the elementary abelian group $\mathbb{Z}_5  \times  \mathbb{Z}_5$ after applying the $\textbf{G1}_{2,3}$(\textbf{W}). 

\[ \textbf{G1}_{2,3}(\textbf{W}) =\left( \begin{array}{ c c c c c }
19 & 4 & 9 & 8 & 11 \\
1 & 14 & 10 & 17 & 15 \\
13 & 3 & 5 & 22 & 2 \\
7 & 23 & 20 & 21 & 16 \\
$*$ & 6 & 12 & 0 & 18 \end{array} \right)\]

\end{example}

\begin{thm}
	The application of the permutation function $\textbf{G1}_{x_1,x_2}$ to a three dimensional periodic Costas array over the elementary abelian group $\mathbb{Z}_p \times \mathbb{Z}_p$ generates a  new three dimensional periodic Costas array over the elementary abelian group $\mathbb{Z}_p  \times  \mathbb{Z}_p$.
\end{thm}

\begin{IEEEproof}
	Let $(i,j), (i', j'), (h_1, h_2) \in \mathbb{Z}_p \times \mathbb{Z}_p$, and $x_1, x_2 \in \{0 \dots p-1\}$, $(x_1, x_2) \ne (0,0)$
	\begin{eqnarray}
		log(x_1i\alpha + x_2j + h_1\alpha + h_2) - log(x_1i\alpha + x_2j)=
	\end{eqnarray}
	
	\begin{eqnarray}
		log \left (\frac{x_1i\alpha + x_2j + h_1\alpha + h_2}{x_1i\alpha + x_2j} \right ) =
	\end{eqnarray}
	
	\begin{eqnarray}
	 log\left (1 + \frac{h_1\alpha + h_2}{x_1i\alpha + x_2j}\right)
	\end{eqnarray}
	
	Proof of the Costas property
	
	\begin{eqnarray}
		 \lefteqn{log\left (1 + \frac{h_1\alpha + h_2}{x_1i\alpha + x_2j}\right)} \nonumber \\
		  & = &   log\left (1 + \frac{h_1\alpha + h_2}{x_1i'\alpha + x_2j'}\right) 
	\end{eqnarray}
	\begin{eqnarray}
		 \lefteqn{\left (1 + \frac{h_1\alpha + h_2}{x_1i\alpha + x_2j}\right)} \nonumber \\
		 & = &\left (1 + \frac{h_1\alpha + h_2}{x_1i'\alpha + x_2j'}\right)
	\end{eqnarray}
	
	Therefore for $(h_1, h_2) \ne (0,0) \rightarrow (i,j) = (i',j')$
	
\end{IEEEproof}

\section{Generalization: $(m+1)$-dimensions multiplication permutation}\label{sec:g1g}

Let  a = $(a_1, a_2, \dots, a_m) \in (\mathbb{Z}_p)^m$, and x = $(x_1, x_2, \dots, x_m) \in (\mathbb{Z}_p)^m$, such that $x_i \in \{1 \dots p-1\}$, and let $(\textbf{w}_{a_1, a_2, \dots, a_m})$ be the elements of $m$ dimensional matrix \textbf{W} in positions $(a_1, a_2, \dots,  a_m)$.

\begin{definition}\textbf{(G1: Multiplication Permutation)}\\
$\textbf{G1}_{x_1, x_2, \dots, x_m}$:$(\mathbb{Z}_p)^m \rightarrow (\mathbb{Z}_p)^m$ is a permutation function such that  $\textbf{G1}_{x_1, x_2, \dots, x_m}$:$(\textbf{w}_{a_1, a_2, \dots, a_m}) \rightarrow (\textbf{w}_{x_1a_1, x_2a_2, \dots, x_ma_m })$ permutes the elements of matrix \textbf{W} from position  $(a_1, a_2, \dots, a_m)$ to position $(x_1a_1, x_2a_2, \dots, x_ma_m)$ of a $(m+1)$ dimensional periodic Costas \textbf{W} over the elementary abelian group $(\mathbb{Z}_p)^m$.
\end{definition}

\begin{thm}
The application of the permutation function \textbf{G1} to a $(m+1)$ dimensional periodic Costas array over the elementary abelian group $(\mathbb{Z}_p)^m$ generates new $(m+1)$ dimensional periodic Costas arrays over the elementary abelian group $(\mathbb{Z}_p)^m$
\end{thm}

\begin{IEEEproof}
 	Let $h, a, b \in GF(p^m)$ and $x \in (\mathbb{Z}_p)^m$, such that $x_i \in \{1 \dots p-1\}$ and let 
	$(x \cdot a) = (x_1a_1\alpha^{m-1} + x_2a_2\alpha^{m-2} + \dots + x_ma_m)$.
	\begin{eqnarray}
		log((a \cdot x) + h) - log(x \cdot a)=\\
		log \left (\frac{(x \cdot a) + h}{x \cdot a} \right ) = log\left (1 + \frac{h}{(x \cdot a)}\right)
	\end{eqnarray}
	
	Proof of the Costas property

	\begin{eqnarray}
		 log\left (1 + \frac{h}{(x \cdot a)}\right) & = & log\left (1 + \frac{h}{(x \cdot b)}\right) \\
		 \left (1 + \frac{h}{(x  \cdot a)}\right) & = &\left (1 + \frac{h}{(x  \cdot b)}\right)
	\end{eqnarray}
	
	Therefore for $h \ne 0 \rightarrow a = b$ 
	
\end{IEEEproof}

\section {Row and Column Shear}\label{sec:g2}

The row (column) shear symmetry consists shifting each of the elements of the row (column) of the two dimensional representation of the 3D Welch MPCA by a different number of shifts. Example \ref{ex:rg2} shows how to obtain a new MPCA by shifting rows 0 through 4 by 0 through 4 times to the right (E.g row 1 is cyclically shifted 1 time to the right, row 2 is cyclically shifted 2 times to the right,  and so on). In the following sections we define the row shear (RG2), column shear (CG2), and the combination of the row and column shear symmetries (G2), and apply them to the 3D Welch MPCA.  Then we define a generalization of the \textbf{G2} symmetry for $(m+1)$ dimensions.

\subsection{Row Shear}

Let $(i, j) \in (\mathbb{Z}_p)^2$, and $x \in \{0 \dots p-1\}$, and let $(\textbf{w}_{i, j})$ be the element of matrix \textbf{W} in position $(i,j)$.

\begin{definition}(Row Shear)\\
$\textbf{RG2}_x: (\mathbb{Z}_p)^2 \rightarrow  (\mathbb{Z}_p)^2$ is a permutation function such that $\textbf{RG2}_x:\textbf{w}_{i, j} \rightarrow \textbf{w}_{i , j+xi}$ shifts the elements of the rows $i$ from position $j$ to position $j+xi$ of a three dimensional periodic Costas \textbf{W} over the elementary abelian group $\mathbb{Z}_p \times \mathbb{Z}_p$.
\end{definition}

\begin{example}\label{ex:rg2}

The following three dimensional periodic Welch Costas array over the elementary abelian group $\mathbb{Z}_5  \times  \mathbb{Z}_5$,  is one symmetry, out of $5$ distinct column stairlike shift symmetries obtained after applying the $\textbf{RG2}_1$(\textbf{W}). 

\[\textbf{RG2}_1(\textbf{W}) = \left( \begin{array}{ c c c c c }
17 & 14 & 15 & 10 & 1 \\
23 & 16 & 20 & 7 & 21 \\
11 & 9 & 19 & 8 & 4 \\
5 & 13 & 22 & 3 & 2 \\
$*$ & 0 & 6 & 18 & 12 \end{array} \right)
\]

\end{example}

\subsection{Column Shear}

Let $(i, j) \in (\mathbb{Z}_p)^2$, and $x \in \{0 \dots p-1\}$, and let $(\textbf{w}_{i, j})$ be the element of matrix \textbf{W} in position $(i,j)$.

\begin{definition}(Column Shear)\\
$\textbf{CG2}_x: (\mathbb{Z}_p)^2 \rightarrow  (\mathbb{Z}_p)^2$ be a permutation function such that $\textbf{CG2}_x:\textbf{w}_{i, j} \rightarrow \textbf{w}_{i + xj, j}$ shifts the elements of the columns $j$ from position $i$ to position $i+xj$ of a three dimensional periodic Costas \textbf{W} over the elementary abelian group $\mathbb{Z}_p \times \mathbb{Z}_p$.
\end{definition}

\begin{example}

The following three dimensional periodic Welch Costas array over the elementary abelian group $\mathbb{Z}_5  \times  \mathbb{Z}_5$,  is one symmetry, out of $5$ distinct column stairlike shift symmetries after applying the $\textbf{CG2}_2(\textbf{W})$. 

\[\textbf{CG2}_2(\textbf{W}) = \left( \begin{array}{ c c c c c }
1 & 8 & 6 & 16 & 5 \\
7 & 22 & 14 & 11 & 12 \\
19 & 0 & 23 & 2 & 10 \\
13 & 17 & 4 & 18 & 20 \\
$*$ & 21 & 3 & 15 & 9 \end{array} \right)
\]
\end{example}

Note that because in the permutation function, the new position in the first dimension depends on the position of the second dimension 
$\textbf{RG2}_{x}(\textbf{CG2}_{y}(\textbf{W})) \ne \textbf{CG2}_{y}(\textbf{RG2}_{x}(\textbf{W}))$.  However the composition of these generators generates another MPCA.

\begin{example}

The following is the three dimensional periodic Welch Costas array over the elementary abelian group $\mathbb{Z}_5  \times  \mathbb{Z}_5$ after applying $\textbf{RG2}_{1}(\textbf{CG2}_{2}(\textbf{W}))$. 

\[  \textbf{RG2}_{1}(\textbf{CG2}_{2}(\textbf{W})) =\left( \begin{array}{ c c c c c }
17 & 9 & 6 & 7 & 2 \\
23 & 13 & 15 & 8 & 12 \\
11 & 0 & 20 & 3 & 1 \\
5 & 14 & 19 & 18 & 21 \\
$*$ & 16 & 22 & 10 & 4 \end{array} \right)
\]

\end{example}

Let $(i, j) \in \mathbb{Z}_p \times \mathbb{Z}_p$, and $x_1, x_2 \in \{0 \dots p-1\}$,  and let $(\textbf{w}_{i, j})$ be the element of matrix \textbf{W} in position $(i,j)$.

\begin{definition}(Row and Column Shear) 
$\textbf{G2}_{x_1,x_2} :(\mathbb{Z}_p)^2 \rightarrow  (\mathbb{Z}_p)^2$ is a permutation function such that $\textbf{G2}_{x_1,x_2}:(\textbf{w}_{i, j}) \rightarrow (\textbf{w}_{i + x_2j , j + x_1i})$ shifts the elements from position $(i,j)$ to position $(i+x_2j, j+x_1i)$ of a three dimensional periodic Costas \textbf{W} over the elementary abelian group $\mathbb{Z}_p \times \mathbb{Z}_p$ whenever $(i+x_2j, j+x_1i) \neq \{0\}$.
\end{definition} 

\begin{thm}\label{thm:shear3d}
	The application of the row and column shift permutation function $\textbf{G2}_{x_1, x_2}$ to a three dimensional periodic Costas array over the elementary abelian group $\mathbb{Z}_p \times \mathbb{Z}_p$ generates a  new three dimensional periodic Costas array over the elementary abelian group $\mathbb{Z}_p \times \mathbb{Z}_p$.
\end{thm}

\begin{IEEEproof}
	Let $(i,j), (i', j'), (h_1, h_2) \in \mathbb{Z}_p \times \mathbb{Z}_p$, and $x_1, x_2 \in \{0 \dots p-1\}$
	\begin{eqnarray}
		\lefteqn{log((i+x_2j)\alpha + j  + x_1i + h_1\alpha + h_2)} \nonumber \\
		& & - log((i+x_2j)\alpha + j + x_1i)=\\
		\lefteqn{log \left (\frac{(i + x_2j)\alpha + j + x_1i + h_1\alpha + h_2}{(i+x_2j)\alpha + j + x_1i} \right )} \nonumber \\
		& & = log\left (1 + \frac{h_1\alpha + h_2}{(i+x_2j)\alpha + j + x_1i}\right)
	\end{eqnarray}
	
	Proof of the Costas property
	
	\begin{eqnarray}
		 \lefteqn{log\left (1 + \frac{h_1\alpha + h_2}{(i+x_2j)\alpha + j + x_1i}\right)} \nonumber \\
		  & &  = log\left (1 + \frac{h_1\alpha + h_2}{(i'+x_2j')\alpha + j' + x_1i'}\right) 
	\end{eqnarray}
	\begin{eqnarray}
		 \lefteqn{ \left (1 + \frac{h_1\alpha + h_2}{(i+x_2j)\alpha + j + x_1i}\right)} \nonumber \\
		 &  & = \left (1 + \frac{h_1\alpha + h_2}{(i'+x_2j')\alpha + j' + x_1i'}\right)
	\end{eqnarray}
	\begin{eqnarray}
		 \lefteqn{ \left (\frac{h_1\alpha + h_2}{(i+x_2j)\alpha + j + x_1i}\right)} \nonumber \\
		 &  & = \left ( \frac{h_1\alpha + h_2}{(i'+x_2j')\alpha + j' + x_1i'}\right)
	\end{eqnarray}
	
	\begin{eqnarray}
		 \lefteqn{ {(i+x_2j)\alpha + j + x_1i}}  \nonumber \\
		 &  & = {(i'+x_2j')\alpha + j' + x_1i'}
	\end{eqnarray}
	\begin{eqnarray}
		\alpha(i-i') + \alpha x_2(j-j') + x_1(i-i')+ (j-j') = 0
	\end{eqnarray}
	
	Therefore for $(h_1, h_2) \ne (0,0) \rightarrow (i,j) = (i',j')$
	
\end{IEEEproof}

\section{Generalization: $(m+1)$-dimensions shear permutation}\label{sec:g2g}

Let  a = $(a_1, a_2, \dots, a_m) \in (\mathbb{Z}_p)^m$, $k,r \in \{1 \dots m\}$, $k \ne r$, and $x \in \{1 \dots p-1\}$, and let $(\textbf{w}_{a_1, a_2, \dots, a_m})$ be the elements of $m$ dimensional matrix \textbf{W} in positions $(a_1, a_2, \dots,  a_m)$.

\begin{definition}\textbf{(G2: Shear Permutation)}\\
$\textbf{G2}$:$(\mathbb{Z}_p)^m \rightarrow (\mathbb{Z}_p)^m$ is a permutation function such that  $\textbf{G2}$:$(\textbf{w}_{a_1, a_2, \dots, a_m}) \rightarrow (\textbf{w}_{ a_1 + a_2u_2 + \dots a_mu_m, a_1v_1 + a_2 + \dots a_mv_m+\dots})$ permutes the elements of matrix \textbf{W} from position  $(a_1, a_2, \dots, a_m)$ to position $(a_1 + a_2u_2 + \dots a_mu_m, a_1v_1 + a_2 + \dots a_mv_m+\dots)$ of a $(m+1)$ dimensional periodic Costas \textbf{W} over the elementary abelian group $(\mathbb{Z}_p)^m$ whenever $(a_1 + a_2u_2 + \dots a_mu_m, a_1v_1 + a_2 + \dots a_mv_m+\dots) \neq \{0\}$.
\end{definition}

\begin{thm}
The application of the permutation function \textbf{G2} to a $(m+1)$ dimensional periodic Costas array over the elementary abelian group $(\mathbb{Z}_p)^m$ generates new $(m+1)$ dimensional periodic Costas arrays over the elementary abelian group $(\mathbb{Z}_p)^m$
\end{thm}

\begin{IEEEproof}
 	The proof is similar to the proof in Theorem \ref{thm:shear3d}
\end{IEEEproof}

\section{Computational results}\label{sec:comp}

The exhaustive exploration for the enumeration of multidimensional periodic Costas arrays (MPCA) has the same computational complexity of the well known problem of enumeration of two-dimensional Costas arrays of size $p^m - 1$, which is $(p^m-1)!$.   More precisely the exhaustive exploration for the enumeration of periodic MPCA over the elementary abelian group is $(p^m-2)!$; because we can fix the first defined (i.e. non zero position) to a constant value $k$ and then permute all the other positions.  This is due to the fact that the arrays are multi-periodic and therefore all multi-periodic shifts are equivalent.   All the other arrays with values different to$k$ are obtained by the addition symmetry.  

At the moment of writing this paper, the last enumeration reported finished for two dimensional Costas arrays is for length 28~\cite{Konstantinos28} and the time per single CPU was determined to be 70 years.  With our resources we were able to enumerate the multidimensional periodic Costas arrays over elementary abelian groups of length $(\mathbb{Z}_3)^2$, $(\mathbb{Z}_5)^2$ and $(\mathbb{Z}_2)^3$, which are similar to computing the two dimensional Costas arrays of lengths 7, 23 and 6 respectively.  The complete enumeration for $(\mathbb{Z}_7)^2$ or $(\mathbb{Z}_3)^3$ would require approximately the equivalent to compute the enumeration of two dimensional Costas arrays of length 47 and  25 which will require high performance computing to be completed in a reasonable amount of time.

\subsection{The algorithm}
The algorithm to compute the multidimensional examples is similar to the algorithm to construct two dimensional Costas arrays~\cite{ArceOrtiz2011,Konstantinos28}.  It consists in a one dimensional permutator that uses a backtracking algorithm to prune the search space.   Backtracking is a general algorithm for solving computational problems by incrementally generating all possible solutions. The execution of a backtracking algorithm can be modelled as a search tree where every node is a partial solution. Moving forward corresponds to approaching a valid solution, and going backwards corresponds to abandoning a partial candidate that cannot possibly generate a valid solution.  In this case the possible solutions are the MPCAs, and the partial solutions are sub permutations that meet the multi-periodic Costas property.  The main change to the original algorithm for two dimensions is in the function to check whether a sub permutation meets the three dimensional or four dimensional Costas property.   We have parallel implementations of these algorithms using MPI, CUDA (GPUs), and FPGAs.

\section{Conclusions}\label{sec:con}

We have presented two generators \textbf{G1} and \textbf{G2} for the group of symmetries of MPCAs over elementary abelian group  $(\mathbb{Z}_p)^m$. By applying \textbf{G1} and \textbf{G2} we characterized the full group of symmetries for the examples $(\mathbb{Z}_3)^2$, $(\mathbb{Z}_5)^2$ and $(\mathbb{Z}_2)^3$.  Because of the strong properties of MPCA, perfect autocorrelation and multi-periodicity, we conjecture that  generators \textbf{G1} and \textbf{G2} characterize the full group of symmetries of the multidimensional periodic Welch Costas arrays over elementary abelian group $(\mathbb{Z}_p)^m$.

We also completed exhaustive search for the cases: $(\mathbb{Z}_4)^2$, $(\mathbb{Z}_2 \times \mathbb{Z}_3)$, $(\mathbb{Z}_3 \times \mathbb{Z}_4)$ with empty results, and for the same reason as before, we conjecture that there are no multidimensional periodic Costas arrays other than multidimensional periodic Costas arrays over the elementary abelian group $(\mathbb{Z}_p)^m$.

 \bibliographystyle{plain}
  \bibliography{references}  


\end{document}